\documentclass[journal,twocolumn,10pt,twoside]{IEEEtran}

\usepackage[T1]{fontenc}
\usepackage[latin9]{inputenc}
\usepackage{amsthm} 
\newtheorem{mydef}{Definition}
\usepackage{graphicx}
\usepackage{setspace,bm}
\usepackage{bbm}
\usepackage{framed,cite}
\usepackage{psfrag}
\usepackage{multicol}
\usepackage{multirow}
\usepackage{widetext} 
\usepackage{blindtext}
\usepackage{amsmath,amssymb} 
\usepackage{color}
\usepackage{tikz}
\usepackage{pgfplots} 
\pgfplotsset{width=7cm}
\usepackage[caption=false,font=footnotesize,subrefformat=parens,labelformat=parens]{subfig}
\usepackage[normalem]{ulem}
\usepackage{algorithm}
\usepackage{algpseudocode}
\usetikzlibrary{shapes.multipart,matrix,positioning}

\newcommand{\varFib}{\mathbf{v}}

\newcommand{\varR}{\mathbf{r}}

\newcommand{\varS}{\mathbf{s}} 
\newcommand{\varTS}{\tilde{\varS}}

\newcommand{\varX}{\mathbf{x}}
\newcommand{\varY}{\mathbf{y}}

\newcommand{\varEta}{\bm{\mu}}
\newcommand{\varSig}{\mathbf{\Sigma}}
\newcommand{\vars}{s}

\newcommand{\varNumSym}{K} 
\newcommand{\spanNum}{N} 
\newcommand{\varNoise}{\mathbf w} 

\newcommand{\varNumPart}{N_p} 

\newcommand{\jlt}{J. Lightw. Technol.}
\newcommand{\ptl}{IEEE Photon. Technol. Lett.}
\newcommand{\ope}{Optics Express}
\newcommand{\tit}{IEEE Trans. Inf. Theory}
\newcommand{\tcom}{IEEE Trans. Commun.}
\newcommand{\ofc}{Proc. Optical Fiber Communication Conference (OFC)}
\newcommand{\oft}{Optical Fiber Technology}
\newcommand{\ecoc}{Proc. European Conference on Optical Communication (ECOC)}
\newcommand{\psst}{Proc. IEEE Photonics Society Summer Topicals}
\newcommand{\icc}{Proc. International Conference on Communications (ICC)}
\newcommand{\twcom}{IEEE Trans. Wireless Commun.}
\newcommand{\jsac}{IEEE J. Sel. Areas Commun.}

\normalsize

\ifCLASSINFOpdf
\else
\fi

\begin{document}
\title{Stochastic Digital Backpropagation with Residual Memory Compensation} 

\author{Naga V. Irukulapati, \IEEEmembership{Student Member, IEEE}, Domenico Marsella,
Pontus Johannisson,  \\ Erik Agrell, \IEEEmembership{Senior Member, IEEE}, Marco Secondini, \IEEEmembership{Member, IEEE}, and Henk Wymeersch, \IEEEmembership{Member, IEEE}
\thanks{N.V. Irukulapati, P. Johannisson, E. Agrell and H. Wymeersch are with the FORCE Research Centre at Chalmers University of Technology, 41296 Gothenburg Sweden (email: vnaga@chalmers.se). D. Marsella and M. Secondini are with the Institute of Communication, Information, and Perception Technologies, Scuola Superiore Sant'Anna, Pisa, Italy. Since Feb. 2015, D. Marsella is with Alcatel-Lucent, Vimercate, Milan, Italy.}
\thanks{This research was supported by the Swedish Research Council (VR) under grant 2013-5642. The simulations were performed in part on resources provided by the Swedish National Infrastructure for Computing (SNIC) at C3SE.

Part of this work was presented at the Optical Fiber Communication Conference (OFC), Los Angeles, CA, Mar. 2015. 

}}

\IEEEspecialpapernotice{(Invited Paper)}


%



\maketitle

\begin{abstract}

Stochastic digital backpropagation (SDBP) is an extension of digital backpropagation (DBP) and  is based on the maximum a posteriori principle. SDBP takes into account noise from the optical amplifiers in addition to handling deterministic linear and nonlinear impairments. The decisions in SDBP are taken on a symbol-by-symbol (SBS) basis, ignoring any residual memory, which may be present due to {non-optimal processing} in SDBP. In this paper, we extend SDBP to account for memory between symbols. In particular, two different methods are proposed: a Viterbi algorithm (VA) and a decision directed approach. Symbol error rate (SER) for memory-based SDBP is significantly lower than the previously proposed SBS-SDBP. For inline dispersion-managed links, the VA-SDBP has up to 10 and 14 times lower SER than DBP  for QPSK and 16-QAM, respectively.

\end{abstract}


%
\IEEEpeerreviewmaketitle

\section{Introduction}\label{secIntro}

\IEEEPARstart{D}{igital} backpropagation (DBP) has  been proposed as a universal technique for jointly compensating for the {intra-channel} linear and nonlinear impairments in the coherent fiber-optic system 
\cite{Essiambre2005,Roberts2006,Ip2010BPSurvey,Li2008AssyBpWdm,Fehenberger2014a}. As a result, the DBP has been used to benchmark schemes proposed in the literature \cite{Koike-Akino2012,Millar2010a,Gao2012,Secondini2014a,Du2014}. The assumed optimality of DBP has spurred intense research in low-complexity variations, including weighted DBP,  perturbation DBP, and filtered DBP \cite{Du2014, Napoli2014d}. {{While the focus of the current paper is on single-channel systems, for wavelength division multiplexing (WDM) systems, DBP is typically employed for the center channel, thereby accounting only for the intra-channel effects. Inter-channel nonlinear effects in WDM systems can be modeled by taking the advantage of the temporal correlations of the nonlinear phase noise using a time-varying system with inter-symbol interference (ISI) and thereby compensating for these inter-channel nonlinear effects \cite{Secondini2014b,Serena2015a,Dar2015}. }}
While DBP has received a great deal of attention, it only deals with deterministic linear and nonlinear impairments and inherently does not consider noise. It is known that the nondeterministic nonlinear effects, such as \emph{nonlinear signal--noise interaction} (NSNI) between the transmitted signal and the amplified spontaneous emission (ASE) noise, limit the transmission performance of a fiber-optic system \cite{Essiambre2010Capacity,Beygi2013,Du2014}. Studies on the impact of NSNI reveal that there is a significant penalty due to NSNI for inline optical dispersion-managed (DM) links, and the severity of the NSNI is dependent on modulation formats and the symbol rate used in the system \cite{Bononi2010nsni,Foursa2013nsni}. It is often argued that NSNI cannot be compensated for in digital signal processing (DSP) due to the nondeterministic nature of ASE noise \cite{Foursa2013nsni} and as a result, none of the DBP methods account for NSNI. To deal with stochastic disturbances, Bayesian detection theory can be used to formulate  maximum a posteriori  probability (MAP) detectors, which are provably optimal in terms of minimizing the error probability.  MAP detectors have been proposed for the discrete memoryless channel \cite{Jiang2011memorylessSDBP} assuming perfect chromatic dispersion (CD) compensation in a DM link, and a look-up table detector that can mitigate data-pattern-dependent nonlinear impairments \cite{Cai2010datadependentMAP}. A low-complexity Viterbi detector is suggested as {an alternative or to complement} DBP for combating fiber nonlinearities \cite{Marsella14jlt}. 
In \cite{Irukulapati2014TCOM}, the stochastic digital backpropagation (SDBP) algorithm was proposed to compensate not only for deterministic linear and nonlinear effects, but also to account for the ASE noise. However, the decisions were taken on a symbol-by-symbol (SBS) basis after applying a matched filter. This approach was later shown to be suboptimal \cite{Irukulapati2014ECOC}.

In this paper, we extend \cite{Irukulapati2014TCOM} to address the sub-optimality in SBS-SDBP by explicitly accounting for residual memory, {which may be present due to matched filtering in SDBP}. In particular, we propose two different methods based on the Viterbi algorithm (VA) and a decision-directed (DD) approach. The VA approach is  similar to \cite{Marsella14jlt},  but does not rely on sending long training sequences to learn the VA branch metrics. The DD approach uses previously  decoded symbols  when taking the decisions for the current symbol and as a result is computationally less complex than the VA approach.
Extensive simulation results indicate significant performance improvements over DBP and SBS-SDBP, in particular for DM links. While the proposed algorithm is computationally complex, we believe this receiver can serve as an inspiration to design low-complexity approaches that still significantly outperform DBP. 

The remainder of this paper is organized as follows.  We first describe the underlying mathematical framework on which SDBP is built, namely  factor graphs (FGs) and message-passing algorithms, in Sec.~\ref{secFG}. In Sec.~\ref{secSectionModel}, the system model is detailed. Sec.~\ref{secApproach} is devoted to the description of  the VA and DD approaches, as well as how these are incorporated into SDBP framework. We present numerical results in Sec.~\ref{secSimulations}, followed by our conclusions in Sec.~\ref{secConclusions}.

\subsubsection*{Notation}
Lower case bold letters (e.g., $\mathbf x$) are used to denote vectors, including sequences of symbols and
 vector representations of continuous-time signals (e.g., through oversampling).
The transpose  of the vector $\varFib$ is denoted by $\varFib^{\text{T}}$. A multivariate Gaussian probability density function (PDF) of a real variable $\varR$ with mean $\varEta$ and covariance matrix $\varSig$ is denoted by $\mathcal{N}(\varR;\varEta,\varSig)$. 

\section{Factor Graphs for Receiver Design}\label{secFG}

Optimal receiver design for digital communications in terms of minimizing the error probability is based on a MAP criterion. However, MAP detectors can be computationally intractable (often with exponential complexity in the dimensionality of the unknown variable), except for certain communication systems. For this reason, much research effort has been devoted in developing \emph{near-MAP detectors}, which can balance near-optimal performance with reasonable computational complexity. A practical framework that has emerged since the early 2000s as a general and automated way to develop near-MAP detectors is that of FGs \cite{Kschischang2001FGSPA,Loeliger2007FG}. An FG is a graph that describes the statistical relation between the variable of interest (i.e., the unknown transmitted data) and the observation (i.e., the received waveform). By performing a message-passing algorithm on such an  FG, it is possible to determine the MAP estimate, or an approximation thereof. 

FGs have been widely used in wireless communication as they provide a methodology to design receivers in a systematic and near-automated way \cite{Wymeersch2007,Worthen2001}. 
{Some applications include message-passing decoders for low-density parity-check (LDPC) and turbo codes \cite{Kschischang1998}, iterative demodulation and decoding for bit-interleaved coded modulation \cite{Caire1998}, joint equalization and decoding \cite{Tuchler2011turboeq}, channel estimation \cite{Guo2009}, timing synchronization \cite{Herzet2007}, and phase-noise recovery \cite{Colavolpe2005}.} It should be noted that while FGs generally lead to the most powerful known receiver algorithms with polynomial complexity, they are often too complex to be implemented as is. For that reason, FGs are often used as a first approach from which practical algorithms can be developed with lower complexity \cite{Tuchler2011turboeq}.

In the context of coherent fiber-optic communication, FGs have only seen limited utilization, mainly due to their high computational complexity. Examples include demodulation \cite{Yu2011}, decoding \cite{Smith2010}, equalization \cite{Djordjevic2008}, and computation of information rates \cite{Djordjevic2005}.  Nevertheless, FGs can serve as a good basis to develop low-complexity receivers. A key application is the design of a near-MAP detector in the nonlinear regime, as proposed in \cite{Irukulapati2014TCOM}. The resulting receiver,  coined SDBP, showed significant performance gains over DBP. {In SDBP, the posterior distribution is obtained by marginalizing the joint distribution of input, all intermediate, unobserved variables in the channel, and the received signal, over the unobserved variables. The statistical relationship between all these variables can be described with a FG (in this case a Markov chain), and the marginalization is performed by message passing. However, the FG-based SDBP receiver proposed in \cite{Irukulapati2014TCOM} is not a true MAP receiver due to a number of heuristic design choices} that were made: (i) a matched filter followed by symbol-rate sampling was employed, similar to DBP, {which is suboptimal and may not give rise to sufficient statistics and may exhibit residual memory; (ii) decisions were made on a SBS basis, ignoring any residual memory;} and (iii) for each SBS decision, a Gaussian approximation  of messages was introduced. The first issue was addressed in \cite{Wymeersch2015SPAWC}, while a possible solution to the third issue was discussed in \cite{Irukulapati2014ECOC}, {considering alternative distributions to a Gaussian in Cartesian coordinates. An alternative approach would be to use a non-parametric approach with a kernel, where the kernel bandwidth is a free parameter that should be tuned \cite{Serdar2011MLOnlyNL}}. The second issue will be addressed in this work.

\section{System Model} \label{secSectionModel}


\begin{figure}
\centering
\psfrag{s}[c][c]{\scriptsize $\varS$}
\psfrag{s'}[c][c]{\scriptsize $\hat \varS$}
\psfrag{st}[c][c]{\scriptsize $\tilde \varS$}
\psfrag{Tx}[c][c]{\scriptsize{Tx}}
\psfrag{SMF}[c][c]{\scriptsize{SMF}}
\psfrag{EDFA1}[c][c]{\scriptsize{EDFA1}}
\psfrag{DCM}[c][c]{\scriptsize{DCM}}
\psfrag{EDFA2}[c][c]{\scriptsize{EDFA2}}
\psfrag{Rx}[c][c]{\scriptsize{Rx}}
\psfrag{Fiber-optical link}[c][c]{\scriptsize{Fiber-optical link}}
\psfrag{N}{{\scriptsize $\times \spanNum$}}
\psfrag{vt1}[c][c]{}
\psfrag{v1}[c][c]{}
\psfrag{rtf}[c][c]{}
\psfrag{r}[c][c]{\vspace{-0.5cm}\scriptsize $\varR$}
\psfrag{Fiber-optical link}[c][c]{{Fiber-optical link}}
\psfrag{for DM link}[c][c]{\scriptsize{for DM link}}
\psfrag{a}{(a)}

\psfrag{pulseshaper}[bl][c][1][90]{\scriptsize{Pulse Shaper}}
\psfrag{SDBP}[bl][c][0.8][90]{\scriptsize{DBP/SDBP}}
\psfrag{Decsn}[bl][c][1][90]{\scriptsize{Decisions}}

\psfrag{M}{\scriptsize{x$M$}}
\psfrag{exp(jGU(t,z))}{\scriptsize $\exp{(j\gamma \Delta |.|^2)}$}
\psfrag{NonLinear}{\scriptsize{Nonlinear}}
\psfrag{Linear}{\scriptsize{Linear}}
\psfrag{HCD}{\scriptsize{$\text{H}_{\text{CD}}$}}
\psfrag{SMF2}[c][c]{\scriptsize{SMF/DCF}}
\psfrag{b}{(b)}

\psfrag{G}{\scriptsize$G_i$}
\psfrag{noise}{$\mathbf w_{ni}$}
\psfrag{EDFAj}{\scriptsize{EDFA1/EDFA2}}
\psfrag{c}{(c)}

\includegraphics[width=1\columnwidth]{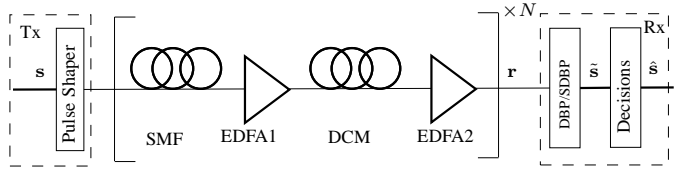}
\caption{\label{fig:FiberLink}\small{A fiber link with $N$ spans where each span consists of an SMF, a DCM module (for DM links), and EDFAs.}}
\end{figure}

{The system that will be considered is a single-channel fiber-optic system as shown in Fig.~\ref{fig:FiberLink}, comprising}  a dual-polarization transmitter block (Tx), including a pulse shaper, a fiber-optic link with $\spanNum$ spans, and a receiver block (Rx) that implements a compensation algorithm followed by a decision unit. Each span of the fiber-optic link consists of a transmission fiber, which is a standard single-mode fiber (SMF) and an optional dispersion-compensating module (DCM) for DM links. In between fiber spans, there are erbium-doped fiber amplifiers (EDFAs) that compensate for the losses in the previous fiber. As indicated in Fig.~\ref{fig:FiberLink}, the transmitted data is denoted by $\varS$, the decoded data by $\hat \varS$, and the received signal by $\varR$. The noise and gain of the EDFAs are resp., denoted by $\varNoise_{ni}$ and $G_i$, where $i\in\{1,2\}$ corresponds to EDFA1 and EDFA2.

A sequence of $\varNumSym$  four-dimensional symbols $\varS=[\vars_1, \vars_2,\ldots ,\vars_\varNumSym]^\text{T} {\in \Omega^{\varNumSym}}$ is transmitted at a symbol rate $1/T_s$ with a pulse-shaping filter $g(t)$, {where  $\Omega \subset \mathbb R^{4}$ is the set of symbols in the four-dimensional constellation, consisting of in-phase and quadrature data from the $x$ and $y$ polarizations. The overall goal of the receiver is to optimally recover $\varS$ from $\varR$. While different optimality criteria can be considered, we aim to minimize the error probability, leading to a MAP receiver, in which the estimate of $\varS$  is 
\begin{align} \label{eqnSeqMAP}
\hat \varS =  \arg \max_{\varS \in \Omega^\varNumSym} p(\varS|\varR),
\end{align}
where $p(\varS|\varR)$ is the a posteriori probability distribution of $\varS$ given the received signal $\varR$. Note that in all derivations, we will consider all signals and vectors to be real.


\section{SDBP and Proposed Approaches} \label{secApproach}

As indicated in Sec.~I, message passing   on an FG relies on local dependencies. In a fiber, the lowest level of local deterministic  dependencies that one can exploit are (i) the linear and nonlinear operation from the split-step Fourier method (SSFM) and (ii) the statistical dependency between input and output of the EDFAs, including ASE noise. Considering the signals after each linear and nonlinear step of each segment of each span of the SSFM \cite[Fig.~1]{Irukulapati2014TCOM}, and the signals after each EDFA, all as \emph{unobserved variables}, one can factorize the joint distribution of the transmitted data $\varS$, the received waveform $\varR$, and these unobserved variables. The posterior distribution in (\ref{eqnSeqMAP}) can thus be interpreted as a marginalization of the joint distribution $p(\varS,\textrm{unobserved variables}|\varR)$. 
This marginalization from the joint distribution to $p(\varS|\varR)$ is done using the framework of FGs and a message-passing algorithm called sum-product algorithm (SPA) \cite{Kschischang1998}. 



\begin{mydef}[Particle representation of a distribution]
{A list of particles (or samples) $\mathbf{x}^{(1)},\ \mathbf{x}^{(2)},\,\ldots,\mathbf{x}^{(\varNumPart)}$, denoted by $\{\mathbf{x}^{(n)}\}_{n=1}^{\varNumPart}$, form a particle representation of a  distribution $p(\mathbf{x})$ when $p(\mathbf{x})\approx 1/{\varNumPart} \sum_{n=1}^{\varNumPart}\delta(\mathbf{x}-\mathbf{x}^{(n)})$.}
\end{mydef}

\begin{mydef}[Distributions associated with particles]
With a list of particles $\{\mathbf{x}^{(n)}\}_{n=1}^{\varNumPart}$ defined over $\mathbb{R}^{4M}$, we associate two distributions: $q_{\mathrm{c}}(\mathbf{x})$ is {a distribution} (obtained, e.g., through a parametric approximation) defined over $\mathbb{R}^{4M}$ for which the particles form a sample representation, while $q_{\mathrm{d}}(\mathbf{x})$ is a distribution defined only over $\Omega^{M}$, with $q_{\mathrm{d}}(\mathbf{x}) \propto q_{\mathrm{c}}(\mathbf{x})$.
\end{mydef}

\subsection{SDBP}


 The main idea of SDBP is to marginalize out the unobserved variables through computing messages, which describe statistically (i.e., in the form of a distribution)  the uncertainty of the corresponding variable. This allows us to obtain a description of $p(\varS|\varR)$. 
The messages are computed backwards, starting with the received signal $\varR$ at span $N$ of the fiber-optic link of Fig.~\ref{fig:FiberLink} until the transmitter is reached.

For the fiber-optic channel, closed-form expressions of the distributions are not possible to derive except for some specific scenarios, so the message/distribution is represented with  a list of $\varNumPart$ particles. These particles are propagated at each stage of the fiber-optic link in Fig.~\ref{fig:FiberLink} starting from $\varR$ as described below.

\begin{algorithm} \hspace{-0.2cm}
\caption{Pseudo-code for implementation of SDBP}
\label{algSDBP}
\begin{algorithmic}[1]
\Procedure{SDBP}{$\varR$}
    \State $\varR^{(n)} \gets \varR \quad \forall n$ \Comment {create $\varNumPart$ replicas of $\varR$}
    \For {$i=N$ to $1$} \Comment Iteration over spans
        \State $\varR^{(n)} \gets (\varR^{(n)}+\varNoise_{n2}^{(n)})/\sqrt{G_2} \quad \forall n$ \Comment {EDFA2} 
          \State $\varR^{(n)} \gets \text{SSFM}_1^{-1}(\varR^{(n)}) \ \quad \quad \quad \forall n$ \Comment {DCM} 
          \State $\varR^{(n)} \gets (\varR^{(n)}+\varNoise_{n1}^{(n)})/\sqrt{G_1}  \quad \forall n$ \Comment {EDFA1} 
          \State $\varR^{(n)} \gets \text{SSFM}_2^{-1}(\varR^{(n)})  \ \quad \quad \quad \forall n$ \Comment {SMF}
\EndFor
        \State {$\varTS^{(n)} \gets \text{MF}(\varR^{(n)}) \quad \forall n$ \Comment {MF followed by sampling at symbol rate}}
\EndProcedure
\end{algorithmic}
\end{algorithm}

We start with the known received waveform $\varR$ in Fig.~1, which exhibits no uncertainty and is thus represented by $N_p$ identical particles (line 2 in Algorithm 1). These particles are passed through the inverse of the EDFA2 block of the last span to get a collection of particles, which describe the  uncertainty regarding the variable before EDFA2 (line 4 in Algorithm 1). The particles are then back propagated through the inverse of the SSFM of the DCM (line 5 in Algorithm 1), where $\text{SSFM}_1^{-1}(\varR^{(n)})$ implements the inverse SSFM for entire fiber span. The particles are then back propagated through EDFA1 (line 6 in Algorithm 1) and through the inverse of the SSFM of the SMF (line 7 in Algorithm 1). This process is repeated for all $\spanNum$ spans. 
Note that when $\varNumPart=1$  and $\varNoise_{n2}^{(n)}=\varNoise_{n1}^{(n)}=0$ for all $n$, these steps are identical to  DBP.

As a final step (line 9 in Algorithm 1), SDBP must compute the message related to the transmitted data $\varS$, based on the message describing the waveform after pulse shaping. A heuristic approach has been used in \cite{Irukulapati2014TCOM}, where
each particle waveform is passed through  a matched filter (MF), matched to the pulse shape, and sampled at the symbol rate\footnote{A matched filter maximizes the signal-to-noise ratio for a signal affected by AWGN noise \cite[Ch. 10]{Lapidoth2009}.} at the optimal sampling times leading to $\varNumPart$ particles, $\{\varTS^{(n)}\}_{n=1}^{\varNumPart}$, with  $\varTS^{(n)} \in \mathbb R^{4\varNumSym}$. 
The particles $\{\varTS^{(n)}\}_{n=1}^{\varNumPart}$ can be viewed as samples from a distribution $q_{\mathrm{c}}(\varS)$ (defined for $\varS \in \mathbb{R}^{4K}$), for which $q_{\mathrm{d}}(\varS)$  (defined for $\varS \in \Omega^{K}$) provides an approximation of {$p(\varS|\varR)$}. 
{It is important to note that $q_{\mathrm{d}}(\varS)$ is only an approximation of $p(\varS|\varR)$ and  need not be identical to $p(\varS|\varR)$}, as the use of a MF followed by sampling at the symbol rate is a heuristic. Hence, performing SBS decisions on the marginals of {$q_{\mathrm{c}}(\varS)$} as in \cite{Irukulapati2014TCOM} may not lead to optimal performance (in terms of minimizing the probability of error, either symbol-wise or sequence-wise). In fact, alternatives to a MF were explored in \cite{Wymeersch2015SPAWC}, indicating  performance  improvements.
{In this paper}, we propose to exploit residual memory {present due to non-optimal processing}, by making a decision  regarding $\varS$ based on the entire distribution $q_{\mathrm{c}}(\varS)$, rather than its marginals, leading to the following detector
\begin{align} \label{eqnAproxSeqMAP}
\hat \varS =  \arg \max_{\varS \in \Omega^\varNumSym} q_{\mathrm{d}}(\varS),
\end{align}
where again $q_{\mathrm{d}}(\varS) \propto q_{\mathrm{c}}(\varS)$.
Solving (\ref{eqnAproxSeqMAP}) is hard for two reasons: (i) the number of possible sequences, $\Omega^{\varNumSym}$, is exponential in $\varNumSym$, making the optimization infeasible for large values of $\varNumSym$, and (ii) for any specific sequence in $\Omega^\varNumSym$, $q_{\mathrm{d}}(\varS)$ is hard to determine since we only have particles $\{\varTS^{(n)}\}_{n=1}^{\varNumPart}$, representing $q_{\mathrm{c}}(\varS)$. In order to address the first issue, we impose a Markov structure onto $q_{\mathrm{d}}(\varS)$. To solve the second issue, the set of particles is smoothed with a distribution, which will be discussed in Sec.~\ref{secVASDBP}. We now present two approaches that use variations of (\ref{eqnAproxSeqMAP}) to make decisions on $\varS$.

\subsection{VA-SDBP}\label{secVASDBP}

Assuming that $q_{\mathrm{d}}(\varS)$ follows a Markov structure with memory\footnote{The memory $L$ is a tuning parameter, where larger $L$ will lead to higher complexity and better performance.} $L \ge 0$, we define $\varX_k=[\vars_{k-1}, \vars_{k-2},\ldots,\vars_{k-L}]^{\mathrm{T}}$ and $\varY_k=[\vars_k \ \varX_k^{\mathrm{T}}]^{\mathrm{T}}$. Then
$q_{\mathrm{d}}(\varS)$ can be factorized as
\begin{align}
\nonumber q_{\mathrm{d}}(\varS) &= \prod_{k=L+1}^\varNumSym q_{\mathrm{d}}(\vars_k|\vars_{k-1}, \vars_{k-2},\ldots,\vars_{k-L}) \\
\label{eqnFactvarTS} &= \prod_{k=L+1}^\varNumSym q_{\mathrm{d}}(\vars_k|\varX_k).
\end{align}
Using Bayes' rule, $q_{\mathrm{d}}(\vars_k|\varX_k)$ can be written as
\begin{align} \label{eqnFactBayes}
q_{\mathrm{d}}(\vars_k|\varX_k)=\frac{q_{\mathrm{d}}(\vars_k,\varX_k)}{q_{\mathrm{d}}(\varX_k)}=\frac{q_{\mathrm{d}}(\varY_k)}{q_{\mathrm{d}}(\varX_k)}.
\end{align}
Using (\ref{eqnFactvarTS}) and (\ref{eqnFactBayes}), (\ref{eqnAproxSeqMAP}) can be rewritten as
\begin{align}
\nonumber \hat \varS &=  \arg \max_{\varS \in \Omega^\varNumSym} \{\ln q_{\mathrm{d}}(\varS)\} \\
\nonumber &=\arg \max_{\varS \in \Omega^\varNumSym} \left \{{\sum_{k=1}^\varNumSym} \left [\ln q_{\mathrm{d}}(\varY_k) - \ln q_{\mathrm{d}}(\varX_k)\right] \right \} \\
\label{eqnLogSeqMAP} &= \arg \min_{\varS \in \Omega^\varNumSym} {\sum_{k=1}^\varNumSym} \psi_k(\vars_k, \varX_k),
\end{align}
where $\psi_k$ is the branch metric and $\varX_k$ is the state used in the VA.

{The values of $q_{\mathrm{d}}(\varY_k)$ and $q_{\mathrm{d}}(\varX_k)$ can be computed by marginalizing $q_{\mathrm{d}}(\varS)$. However, we have access only to $q_{\mathrm{c}}(\varS)$ through the particles $\{\varTS^{(n)}\}_{n=1}^{\varNumPart}$. Denoting the appropriate sub-sequences from $\varTS^{(n)}$ by $\varY_k^{(n)}$ and $\varX_k^{(n)}$, we find that  $q_{\mathrm{c}}(\varY_k)\approx 1/{\varNumPart} \sum_{n=1}^{\varNumPart}\delta(\varY_k-\varY_k^{(n)})$ and $q_{\mathrm{c}}(\varX_k)\approx 1/{\varNumPart} \sum_{n=1}^{\varNumPart}\delta(\varX_k-\varX_k^{(n)})$. In order to evaluate
$q_{\mathrm{d}}(\varY_k)$ and $q_{\mathrm{d}}(\varX_k)$, we can impose a parametric approximation for $q_{\mathrm{c}}(\varY_k)$ and $q_{\mathrm{c}}(\varX_k)$ for which the logarithm is easy to compute.} The Gaussian distribution is such a parametric approximation.\footnote{{A non-parametric approach with a kernel can be used as an alternative \cite{Serdar2011MLOnlyNL}, where the kernel bandwidth is a free parameter that should be tuned.}} Hence $q_{\mathrm{c}}(\varY_k)= \mathcal N(\varY_k;\varEta^y_k,\varSig^y_k)$ and $q_{\mathrm{c}}(\varX_k)= \mathcal N(\varX_k;\varEta^x_k,\varSig^x_k)$. The means $\varEta^y_k$, $\varEta^x_k$ and covariances $\varSig^y_k$, $\varSig^x_k$ are estimated as
\begin{align} \label{eqnMeanCovyx}
\nonumber\varEta^y_k=\frac{1}{\varNumPart} \sum_{n=1}^{\varNumPart} {\varY_k^{(n)}} , \quad \varEta^x_k=\frac{1}{\varNumPart} \sum_{n=1}^{\varNumPart} {\varX_k^{(n)}}\\ \nonumber \varSig^y_k=\frac{1}{\varNumPart-1} \sum_{n=1}^{\varNumPart} (\varY_k^{(n)}-\varEta^y_k)(\varY_k^{(n)}-\varEta^y_k)^{\mathrm{T}}, \\
\varSig^x_k=\frac{1}{\varNumPart-1} \sum_{n=1}^{\varNumPart} (\varX_k^{(n)}-\varEta^x_k)(\varX_k^{(n)}-\varEta^x_k)^{\mathrm{T}}.
\end{align}
The factors in (\ref{eqnFactBayes}) can be written as
\begin{align}
q_{\mathrm{d}}(\varY_k) &\propto {\exp\left\{ -\frac{1}{2}\left(\varY_{k}-\varEta^{y}_{k}\right)^{\mathrm{T}}(\varSig^{y}_{k})^{-1}\left(\varY_{k}-\varEta^{y}_{k}\right)\right\} }, \label{eqnQy}\\
q_{\mathrm{d}}(\varX_k) &\propto {\exp\left\{ -\frac{1}{2}\left(\varX_{k}-\varEta^{x}_{k}\right)^{\mathrm{T}}(\varSig^{x}_{k})^{-1}\left(\varX_{k}-\varEta^{x}_{k}\right)\right\} }.\label{eqnQx}
\end{align}
As a result, $\psi_k$ can be simplified as
\begin{align}
\nonumber\psi_k(\vars_k, \varX_k)\propto & \left(\varY_{k}-\varEta^{y}_{k}\right)^{\mathrm{T}}(\varSig^{y}_{k})^{-1}\left(\varY_{k}-\varEta^{y}_{k}\right)\\
\label{eqnMetric}&-\left(\varX_{k}-\varEta^{x}_{k}\right)^{\mathrm{T}}(\varSig^{x}_{k})^{-1}\left(\varX_{k}-\varEta^{x}_{k}\right).
\end{align}
To find an estimate $\hat \varS$ using (\ref{eqnLogSeqMAP}), a VA\footnote{We assume a uniform a priori distribution over all the states which implies the symbols at the start of trellis are unknown. } is used with the current state as $\varX_k$  and the current symbol as $\vars_k$ with branch metric as in (\ref{eqnMetric}) for the $k$th symbol slot. {Observe that since the search space for $\varS$ is $\Omega ^{\varNumSym}$, the search space for  $\varY_k$ is $\Omega^{L+1}$ and $\varX_k \in \Omega^{L}$. When $L=0$, VA-SDBP reverts back to SBS-SDBP from \cite{Irukulapati2014TCOM}.}

\subsection{DD-SDBP}\label{sec:DD-SDBP}

The second approach, DD-SDBP, combines the {idea of exploiting memory, as in VA-SDBP, with  taking decisions on a SBS basis, as in SBS-SDBP.} In this approach, the previously decoded symbols, $[\hat \vars_{k-1}, \hat \vars_{k-2},\ldots, \hat \vars_{k-L}]^\text{T}$, are used while taking decisions for the current symbol $\varS_k$  and as a result  $\varX_k$ in (\ref{eqnMetric}) can be interpreted as a constant that does not affect the optimization in (\ref{eqnMetric}). 
Thus, decisions on $\vars_k$ in DD-SDBP are taken as
\begin{align} \label{eqnDDSDBP}
\nonumber \hat \vars_k  & =  \arg \max_{\vars_k \in \Omega} q_{\mathrm{d}}(\vars_{k}|\hat \vars_{k-1}, \hat \vars_{k-2},\ldots, \hat \vars_{k-L})\\
& =  \arg \max_{\vars_k \in \Omega} \psi (\vars_{k}, \hat \varX_{k}),
\end{align}
which can be solved recursively starting\footnote{For $k=1,\,\ldots,L$, decisions on $\vars_k$ are taken on an SBS basis.} from $k=L+1$ using a similar Gaussian approximation as in VA-SDBP. {However, in contrast to VA-SDBP, DD-SDBP can only account for causal memory effects. }
Note that the search space for $\varY_k$ of (\ref{eqnQy}) is $\Omega$ instead of $\Omega^{L+1}$ as the decisions have to be taken only for $\vars_k$.  When $L=0$,  DD-SDBP also reverts to SBS-SDBP.


\section{Numerical Simulations and Discussion} \label{secSimulations}

\subsection{Simulation Setup and Performance Metrics}


The simulation setup is shown in Fig.~\ref{fig:FiberLink}. The pulse shape used at the transmitter is a root-raised-cosine pulse with a roll-off factor of 0.25 and truncation length of 16 symbol periods. The simulations are performed for a polarization-multiplexed signal, {with no polarization mode dispersion\footnote{{Effect of PMD on DBP and SBS-SDBP has been reported in \cite{Irukulapati2014TCOM}.}}}, either with 16-QAM or QPSK as modulation format and symbol rates $R_s$ of 14 Gbaud, 28 Gbaud, and 56 Gbaud. In each polarization, {$K=4096$} symbols\footnote{{Except for 56 GBd NDM links, where we have simulated with $K=8192$ symbols to properly account for the memory in the system.}} are transmitted in each block of the Monte Carlo simulation. This signal is input to the channel with $N$ spans. The parameters used for the SMF are {$D=16$ ps/(nm km), $\gamma=1.3~\text{(W km)}^{-1}$, $\alpha=0.2$ dB/km,} which are according to the ITU-T G.652. {We have considered  a fiber Bragg grating (FBG) as a DCM.\footnote{{However, a dispersion compensating fiber (DCF), simulated according to G.655 specifications, exhibited similar trends as the FBG.}} }Propagation in the fibers is simulated using the SSFM with a segment length \cite{Zhang2008stepSize} of $\Delta = (\epsilon L_\text{N} L_\text{D}^{2})^{1/3},$ where $\epsilon  = 10^{-4}$, $L_\text{N}=1/(\gamma P)$ is the nonlinear length, $L_\text{D} = {T^{2} 2 \pi c}/(|D|\lambda^2)$ is the dispersion length, $\lambda$ is the wavelength, $c$ is the speed of the light, and $P$ is the average input power to each fiber span. 
We used the same segment lengths for simulating the channel and for both DBP and SDBP.

\begin{table}
\begin{centering}
%

\bigskip

\caption{\label{tabNumSpans}Number of spans, $N$, used in DM and NDM links}
\scalebox{1}{
\begin{tabular}{|c|c|c|c|}
\hline
 & $R_s$ [GBd] & DM & NDM\tabularnewline
\hline
\hline
\multirow{3}{*}{QPSK} & 14 & $50$ & $110$\tabularnewline
\cline{2-4}
 & 28 & $35$ & $110$\tabularnewline
\cline{2-4}
 & 56 & $35$ & $110$\tabularnewline
\hline
\multirow{3}{*}{16-QAM} & 14 & $50$ & $110$\tabularnewline
\cline{2-4}
 & 28 & $40$ & $110$\tabularnewline
\cline{2-4}
 & 56 & $40$ & $110$\tabularnewline
\hline
\end{tabular}
}

\par\end{centering}
\end{table}

The number of spans, $N$, used in each of the scenarios is summarized in Table \ref{tabNumSpans}. 
We also considered a non-DM (NDM) system, wherein there are no DCMs. The span length used for SMF, $L_\textrm{SMF}$, is $80$ km for 16-QAM and $120$ km for QPSK.\footnote{The number of spans and span lengths are selected such that the symbol error rate for DBP is less than $0.001$.} 
FBG with an insertion loss of 3 dB and perfect dispersion compensation for the preceding SMF {is used}. The launch power into the DCM is 4 dB below that of the transmission fiber, which is compensated for by the EDFA after DCM. The noise figure is 5 dB for each of the amplifiers. Ideal low-pass filters with one-sided bandwidth of $R_s$ are used in EDFAs and in the beginning of the receiver. The filtered signal is sent to DBP and three different SDBP detectors:
\begin{enumerate}
\item SBS-SDBP from \cite{Irukulapati2014TCOM};
\item DD-SDBP for $L \in \{ 1, 2\}$, as proposed in Sec.~\ref{sec:DD-SDBP}; and
\item VA-SDBP for $L \in \{ 1, 2\}$, as proposed in Sec.~\ref{secVASDBP}. 
\end{enumerate}
In all SDBP detectors, $N_p=500$ particles were used to generate the results, but we verified that even with $N_p=1500$, similar performance was obtained.  
The receiver is assumed to have perfect knowledge of the polarization state, as well as the carrier phase and the symbol timing. 

We consider two performance metrics. To capture the absolute performance of each detector, we determine the symbol error rate (SER). To capture the relative performance gain of SDBP over DBP, we introduce
$G_{\text{X}} = \mathrm{SER}_{\text{DBP}}/\mathrm{SER}_{\text{X-SDBP}}$, where $\mathrm{X} \in  \{ \mathrm{SBS}, \mathrm{DD},\mathrm{VA}\}$, and in  which  $\mathrm{SER}_{\text{DBP}}$ and $\mathrm{SER}_{\text{X-SDBP}}$ are \emph{lowest} SERs obtained for the respective algorithms. 

\subsection{Results and Discussion}

The SER as a function of input power is shown in Fig.~\ref{fig:SER_Pin_FBG_56G} for the 56 Gbaud system with FBG as DCM for (a) QPSK and (b) 16-QAM. Due to complexity reasons, VA-SDBP is simulated only with $L=1$ for 16-QAM.  Comparing the SER of SBS-SDBP with DD-SDBP/VA-SDBP, we can conclude that by taking the residual memory into account, the SER is significantly reduced. One can also see that for both VA-SDBP and DD-SDBP, increasing $L$ leads to a decrease in the SER. We expect this gain to saturate as $L$ increases. We also see that the optimal power (i.e., corresponding to the lowest SER) varies for each detector: compared to DBP, the optimal power is up to 2 dB larger for QPSK and up to 4 dB larger for 16-QAM.


\begin{figure}[h]
\centering
  \mbox{\subfloat[]{\label{fig:SER_SMF_FBG_QPSK_56G} \includegraphics[width=1\linewidth]{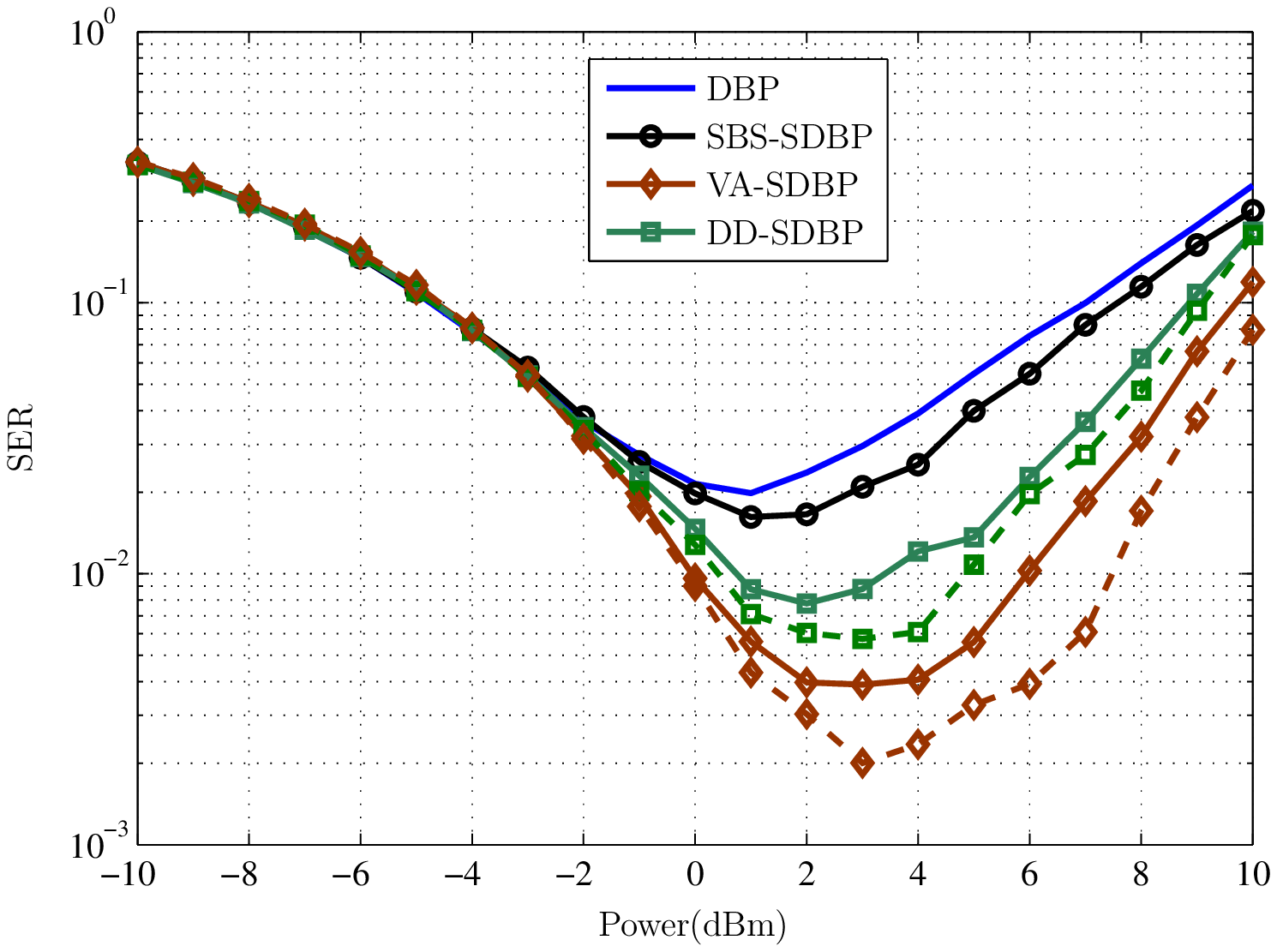}}}
  \mbox{\subfloat[]{\label{fig:SER_SMF_FBG_16QAM_56G} \includegraphics[width=1\linewidth]{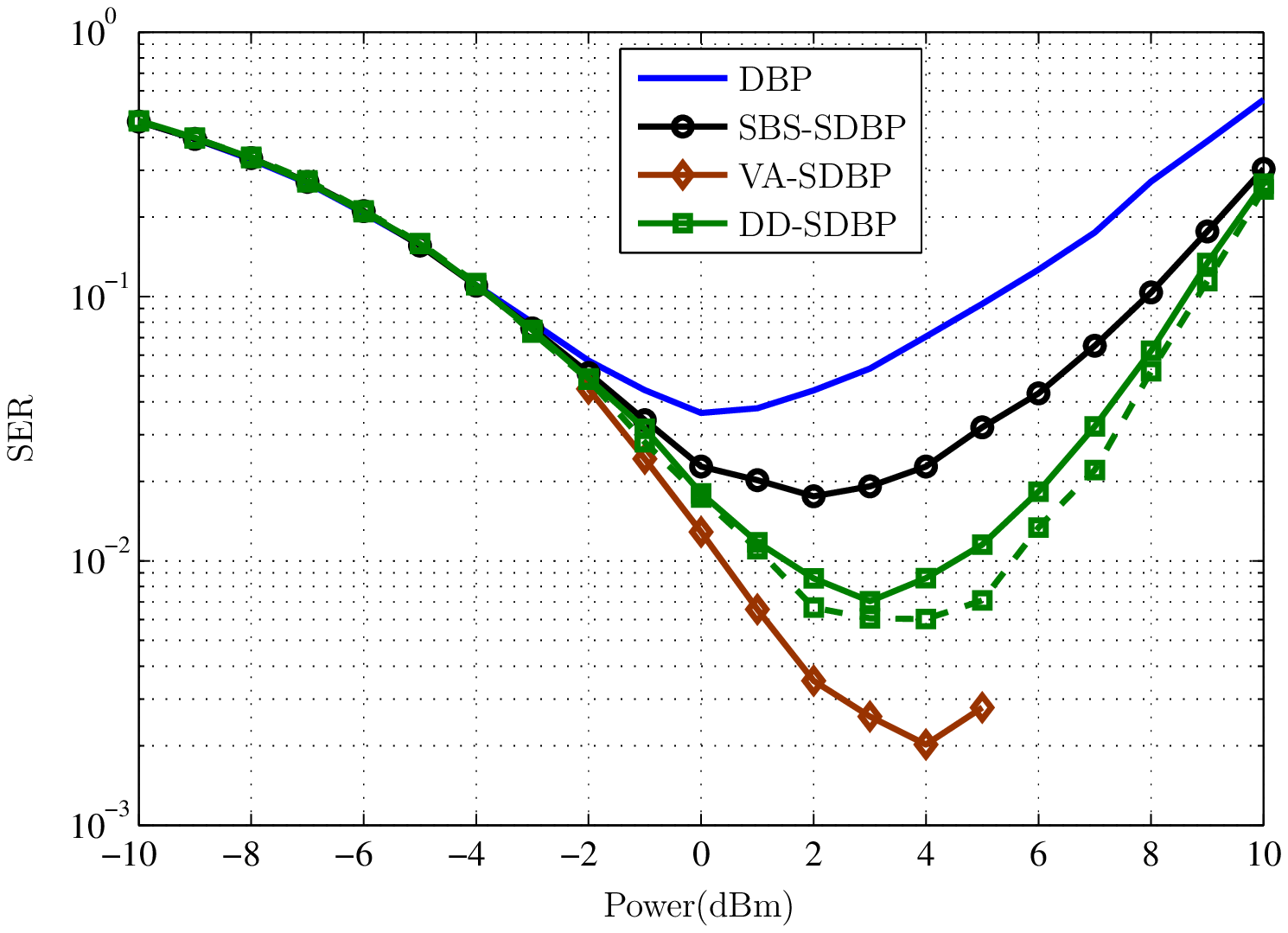}}}
  \caption{\small{{SER as a function of input power for 56 Gbaud, FBG as DCM, for (a) QPSK and (b) 16-QAM. Solid (resp. dashed) lines in VA-SDBP and DD-SDBP represent cases when $L=1$ (resp. $L=2$).}}}
  \label{fig:SER_Pin_FBG_56G}
\end{figure}

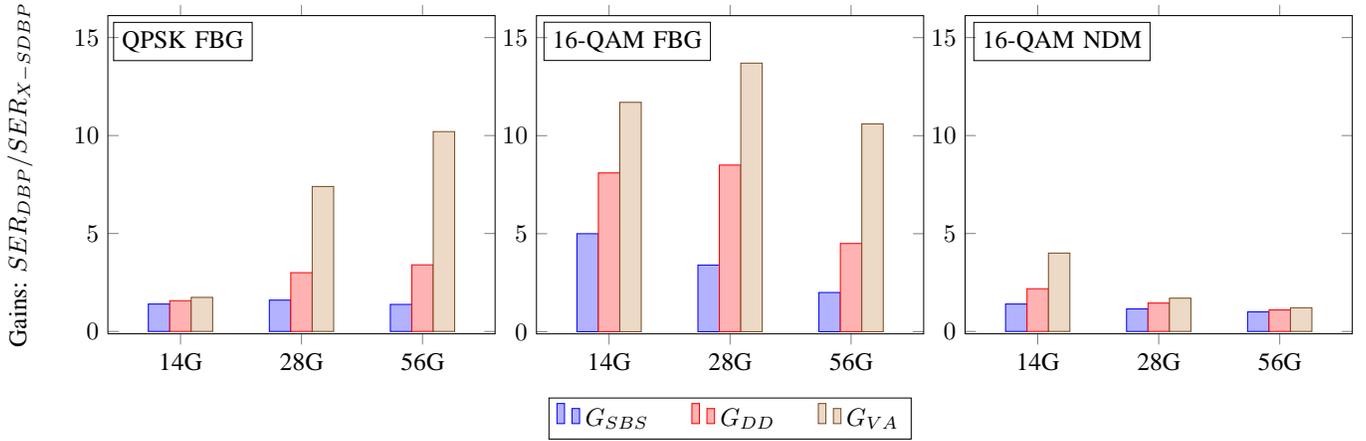
\begin{figure*}
\begin{tikzpicture}[thick,scale=0.95]

\begin{axis}[
    ybar=0pt,
    enlargelimits=0.08,
    bar width=0.3cm,
    ymin=1,ymax=15,
    enlarge x limits=0.3, 
    legend style={at={(0.5,-0.2)},anchor=north,legend columns=-1},	
    xtick=data, 
	ylabel={Gains: $SER_{DBP}/SER_{X-SDBP}$},
    symbolic x coords={14G, 28G, 56G},
    legend style={/tikz/every even column/.append style={column sep=15pt}}
]
\addplot
	coordinates {(14G, 1.4) (28G, 1.6) (56G,1.38)};
\addplot
	coordinates {(14G, 1.57) (28G, 3) (56G,3.4)};
\addplot
	coordinates {(14G, 1.74) (28G, 7.4) (56G,10.2)};
\node[draw] at (2,138) {QPSK FBG};
\end{axis}

\hspace{6cm}

\begin{axis}[
    ybar=0pt,
    enlargelimits=0.08,
    bar width=0.3cm,
    ymin=1.01,ymax=15,
    enlarge x limits=0.3, 
    legend style={at={(0.5,-0.2)},anchor=north,legend columns=-1},	
    xtick=data, 
    symbolic x coords={14G, 28G, 56G},
    legend style={/tikz/every even column/.append style={column sep=15pt}}
]
\addplot
	coordinates {(14G, 5) (28G, 3.4) (56G,2)};
\addplot
	coordinates {(14G, 8.1) (28G, 8.5) (56G,4.5)};
\addplot
	coordinates {(14G, 11.7) (28G, 13.7) (56G,10.6)};
\node[draw] at (15,138) {16-QAM FBG};
\legend{$G_{SBS}$, $G_{DD}$,$G_{VA}$}
\end{axis}

\hspace{6cm}

\begin{axis}[
    ybar=0pt,
    enlargelimits=0.08,
    bar width=0.3cm,
    ymin=1,ymax=15,
    enlarge x limits=0.3, 
    legend style={at={(0.5,-0.1)},anchor=north,legend columns=-1},	
    xtick=data, 
    symbolic x coords={14G, 28G, 56G},
]
\addplot
	coordinates {(14G, 1.4) (28G, 1.15) (56G,1)};
\addplot
	coordinates {(14G, 2.18) (28G, 1.45) (56G,1.1)};
\addplot
	coordinates {(14G,4) (28G,1.7) (56G,1.21)};
\node[draw] at (20,138) {16-QAM NDM};
\end{axis}

\end{tikzpicture}

\caption{\label{figQPSK16QAMGains}Gains in SER for the proposed algorithms compared to DBP.} 
\end{figure*}

Similar behavior is observed for other symbol rates, although we do not show all results. Instead, a summary of the performance gains is presented in Fig.~\ref{figQPSK16QAMGains} for QPSK {with $L=2$} and for 16-QAM {with $L=1$. As the complexity of VA-SDBP grows exponentially with $L$, the same value of $L$ was used for both DD-SDBP and VA-SDBP to have a fair comparison. 
From Fig.~\ref{figQPSK16QAMGains}, irrespective of the symbol rate, DM or NDM links, we observe a clear trend: $\textrm{SER}_{\textrm{VA-SDBP}}< \textrm{SER}_{\textrm{DD-SDBP}} < \textrm{SER}_{\textrm{SBS-SDBP}} < \textrm{SER}_{\textrm{DBP}}$.
{The VA-SDBP can account for both causal and non-causal effect, giving it a performance benefit over DD-SDBP. SBS-SDBP ignores both causal and non-causal memory and thus exhibits the worst performance among the three selected approaches.}
{The decreasing gains with increasing symbol rate for SBS-SDBP can be explained as follows. The larger the deviation of the particle clouds, given by $\{\varTS^{(n)}\}_{n=1}^{\varNumPart}$, from a circular symmetric Gaussian, the higher are the expected gains in SDBP compared to DBP. As the symbol rate increases, the particle clouds in SBS-SDBP tend to become more circular Gaussian and hence the gains decrease. Also for a DM link, we have observed that the particle clouds are less circular Gaussian and hence gains are higher for SDBP in DM links compared to NDM links.}}}


{The gains in VA-SDBP for QPSK increase with increasing symbol rate whereas for 16-QAM, the gains increase from 14 GBd to 28 GBd and then the gains decrease. This maybe due to the use of $L=1$ for 16-QAM, which is not sufficient to account for the residual memory, especially at high symbol rates.}
The main drawback of VA-SDBP is its complexity, which grows exponentially with the memory $L$.  So, an interesting case would be to test a low-complexity version of VA for 16-QAM with higher memory. DD-SDBP is a tradeoff between SBS-SDBP and VA-SDBP, in terms of complexity and performance. Irrespective of which SDBP approach is used, there is always an improvement in terms of SER compared to the traditional DBP algorithm (i.e., $G_{\text{X}} > 1$). This means that NSNI plays an important role in the systems under consideration and one can gain significantly by taking these interactions into account. The NSNI is more  important in the DM systems than the NDM systems, so that gains are lower in NDM systems. An additional observation that can be made from Fig.~\ref{figQPSK16QAMGains} is that the gains are in general higher for 16-QAM than for QPSK {as 16-QAM has more nonlinearities than QPSK and hence more signal-noise interactions, and thereby more gains of SDBP approach compared to DBP}. {Gains in QPSK NDM links (not shown here) turn out to be lower than corresponding gains of 16-QAM NDM case.}

\section{Conclusions}\label{secConclusions}
We have extended the SDBP algorithm to account for residual memory that may be present due to {non-optimal processing in SDBP}. Specifically, we proposed DD-SDBP and VA-SDBP to account for this memory, at an increased cost in terms of complexity. Extensive simulations were performed to evaluate these methods for 16-QAM and QPSK,  and for DM and NDM links.  Results suggest a significant improvement by the proposed detectors for DM links with up to 10 times lower SER for QPSK and up to 14 times lower SER for 16-QAM, compared to DBP. 

{The VA-SDBP can provide optimal decisions on the transmitted sequence, but does so at a high computational cost. Alternatives to consider are low-complexity variations of the VA, as well as algorithms that provide symbol-wise optimal decisions, such as the Bahl-Cocke-Jelinek-Raviv (BCJR) algorithm \cite{Bahl1974}. }

%

Further gains over the proposed algorithms  may be possible and {remains the topic} of ongoing and future research. Lower SER can be expected by increasing the memory in VA-SDBP until the SER gains saturate. In addition, the use of a matched filter is not necessarily optimal.
Initial results in this direction can be found in \cite{Wymeersch2015SPAWC}. Finally, in SBS-SDBP, DD-SDBP, and VA-SDBP, the particles after matched filtering and sampling are approximated with a multivariate Gaussian distribution, which need not be a good approximation, especially at high input powers. Other types of distributions should be considered. 

\bibliographystyle{IEEEtran}

\end{document}